\documentclass{elsart}

\usepackage{graphicx}

\newcommand\HF{{\mathrm{HF}}}
\newcommand\HL{{\mathrm{HL}}}

\begin{document}
\runauthor{Lombardo, Grassi, Forte, Angilella, Pucci, March}
\runtitle{Electronic structure of clusters (LiBC)$_n$}
\journal{Phys. Lett. A}
\received{\today}

\begin{frontmatter}
\title{Electronic structure of clusters (LiBC)$_n$:\\
   $n=1$, 2 and 4}
\author[Chem]{G. M. Lombardo,}
\author[Chem]{A. Grassi,}
\author[Chem]{G. Forte,}
\author[Phys]{G. G. N. Angilella,}
\author[Phys]{R. Pucci,}
\and
\author[RUCA,Oxford]{N. H. March}
\address[Chem]{Dipartimento di Scienze Chimiche, Facolt\`a di Farmacia,
   Universit\`a di Catania,\\ 
Viale A. Doria, 6, I-95126 Catania, Italy}
\address[Phys]{Dipartimento di Fisica e Astronomia, Universit\`a di
   Catania, and\\
Istituto Nazionale per la Fisica della Materia, UdR Catania,\\
Via S. Sofia, 64, I-95123 Catania, Italy}
\address[RUCA]{Department of Physics, University of Antwerp,\\
Groenenborgerlaan 171, B-2020 Antwerp, Belgium}
\address[Oxford]{Oxford University, Oxford, UK}

\maketitle

\begin{abstract}%
A crystalline form of LiBC is known which has been predicted to be
   superconducting, with a $T_c$ comparable to that of MgB$_2$.
In both compounds, superconductivity is enhanced by the presence of
   two electronic bands, one of which is close to a dimensional
   crossover.
Here, we take a quantum chemical approach, and investigate the
   structural and electronic properties of small clusters (LiBC)$_n$
   ($n=1-4$).
With increasing cluster size, we find that several properties of
   crystalline LiBC evolve naturally from the corresponding properties
   of the clusters.
In particular, one may recognize the origin of the solid bilayered
   structure, typical of magnesium diboride, and the character of the
   electronic $\sigma$-band, arising from the overlap of the atomic
   orbitals in the in-plane BC rings. 
Two aspects especially emphasized are \emph{(a)} the HOMO-LUMO gap as
   function of $n$ and \emph{(b)} the role of different spin
   multiplicity.\\
\medskip
\noindent {\sl PACS:}
36.40.Mr, 
36.40.Cg,
74.70.Ad 
\end{abstract}
\end{frontmatter}

A crystalline form of hole-doped LiBC has been studied, which has been
   predicted to be superconducting with a transition temperature $T_c$
   comparable to that of the isoelectronic compound MgB$_2$
   \cite{Rosner:02,Dewhurst:03}.
The structure of LiBC is closely related to the bilayered structure of
   MgB$_2$, with Li replacing Mg, and B$_2$ being replaced by BC, but
   with hexagonal BC layers alternating so that B is nearest neighbour
   to C both within the in-plane rings, and along the $c$~axis.
Although the number of valence electrons decreases by one in replacing
   Mg with Li, this is compensated by the substitution of B$_2$ by BC.
At variance with its similarities with MgB$_2$, a distinctive feature
   of LiBC is that the Li content in Li$_x$BC can be varied with
   respect to its stoichiometric value $x=1$, without any appreciable
   change of crystalline structure for a quite wide range in $x=0.24-1$
   \cite{Rosner:02}, thus allowing to study the superconducting
   properties of this material upon self-hole-doping.

Both MgB$_2$ and LiBC are characterized by a similar electronic
   structure, with a three-dimensional (3D) $\sigma$ subband, mainly
   arising from the B$_2$ (respectively, BC) layers, and a
   quasi-bi-dimensional (quasi-2D) $\pi$ subband, mainly arising from the
   electrons delocalization across the layers.
The relevance of the proximity of the $\pi$ subband to a 3D--2D
   crossover, \emph{i.e.} an electronic topological transition, for
   the dependence of $T_c$ and of the isotope exponent on doping has
   been emphasized elsewhere within the two-band model of
   superconductivity \cite{Angilella:04i}.

Thus, it would be natural to think of both structural and electronic
   properties of crystalline LiBC as arising from the underlying structural and
   electronic properties of the individual LiBC units.
This has motivated the present study of the electronic structure of
   the small clusters (LiBC)$_n$, with $n$ going from 1 to 4.
With increasing cluster size $n$, the solid crystalline and electronic
   structure of LiBC can be approached, thus allowing one to recognize
   the origin of the peculiar bilayered structure of the diborides,
   and the nature of their two coupled electronic bands.
The present study shows that such features are already present in
   clusters (LiBC)$_n$, with $n$ as small as 4.

For these clusters the electronic stuctures, optimized geometries, and
   the vibrational analysis were obtained using the {\sc Gaussian 03}
   package \cite{Frisch:03}, with \emph{ab initio} self-consistent 
   field Hartree-Fock wavefunctions at the 6-311+G(d) level of the
   theory (including polarization and diffuse functions).
Each cluster was studied with various spin multiplicities depending on
   the number of possible uncoupled valence electrons.
Here, we report only those which attained a true minimum.

\begin{figure}[t]
\centering
\begin{tabular}{ccc}
\includegraphics[width=0.3\columnwidth]{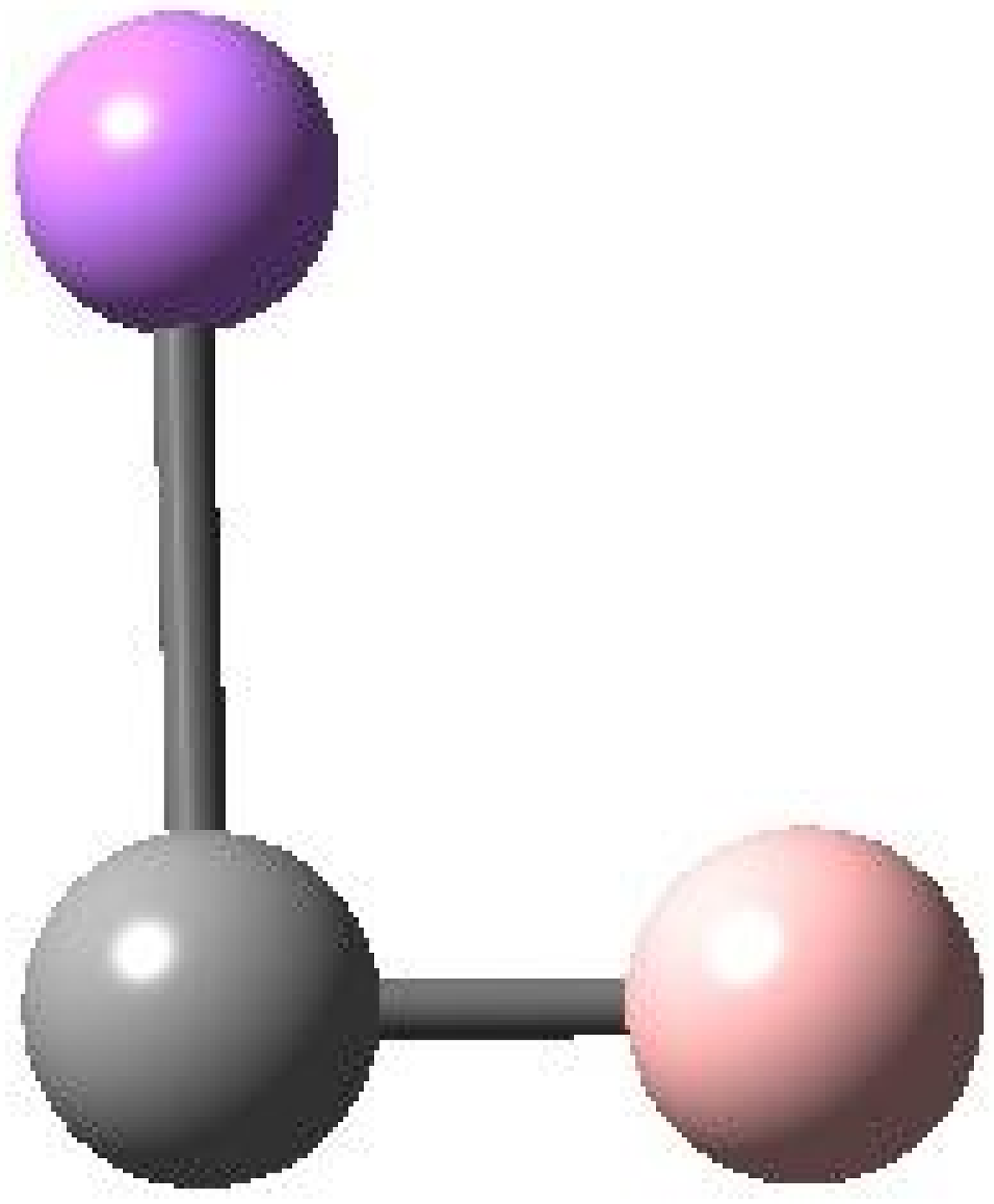}
   &
\includegraphics[width=0.3\columnwidth]{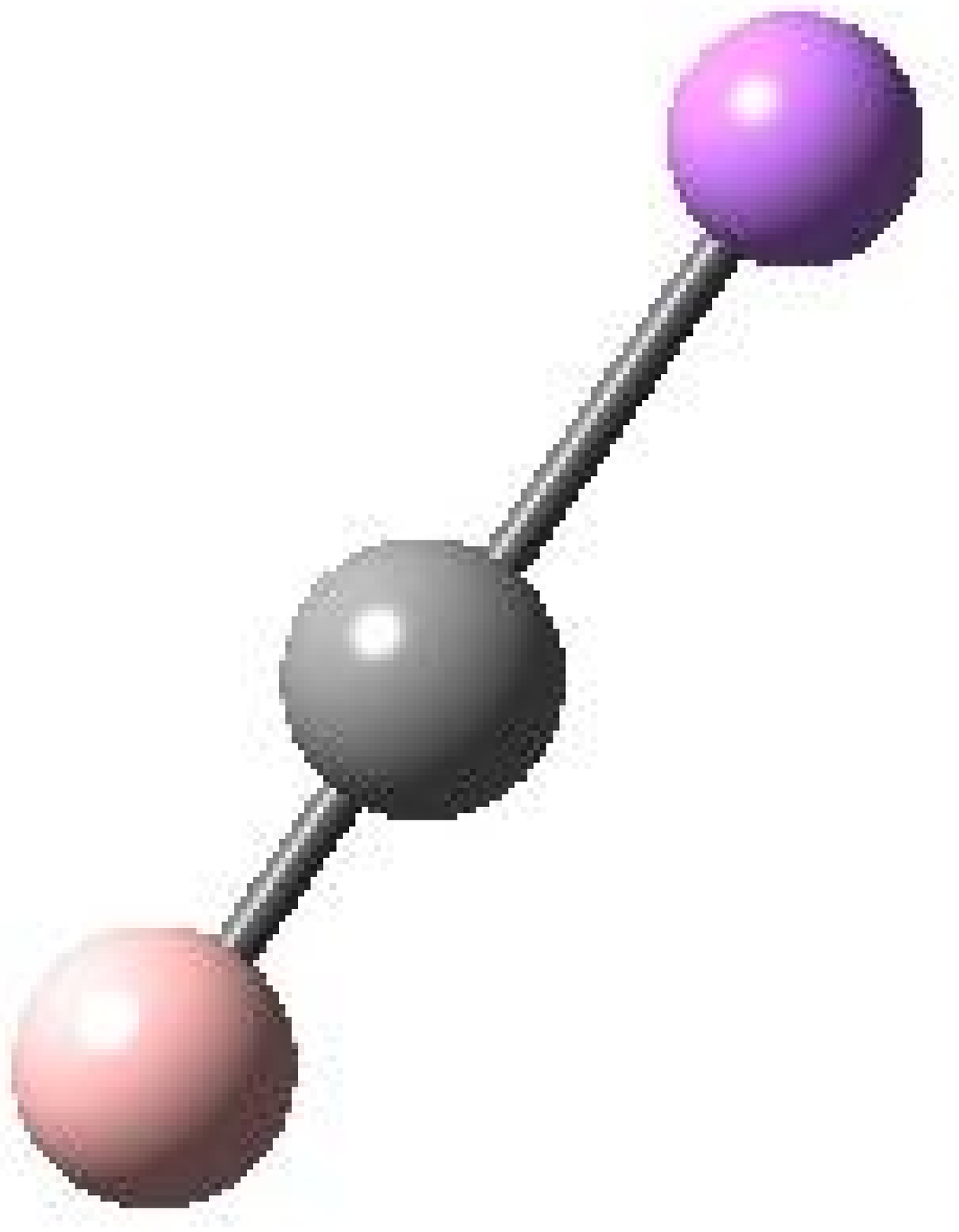}
   &
\includegraphics[width=0.3\columnwidth]{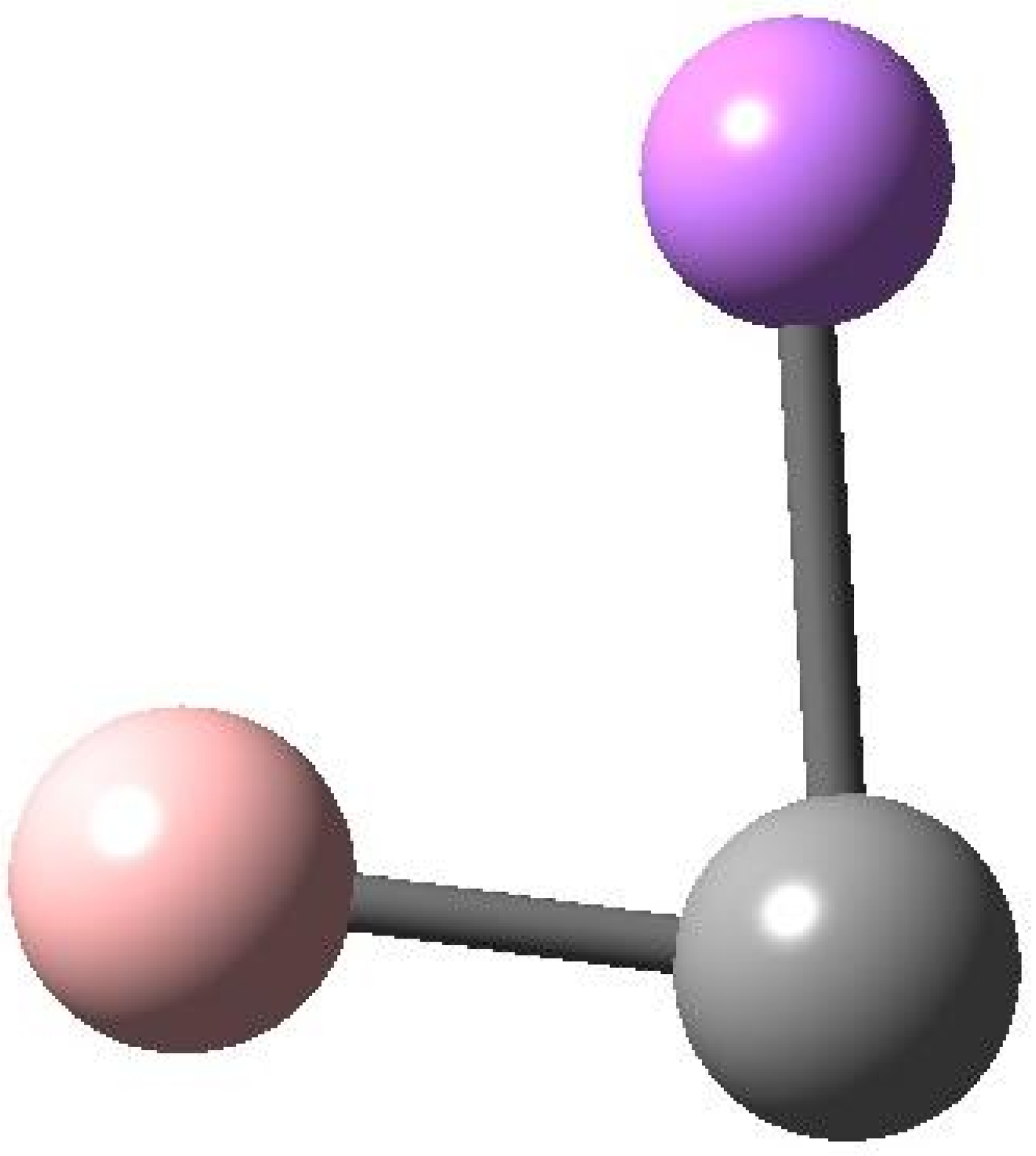}
   \\
$S=1$ & $S=3$ & $S=5$
\end{tabular}
\caption{Fully optimized configuration of the LiBC singlet cluster for
   $S=1$, $3$, $5$ (left to right; Li: purple; B: pink; C: grey).
}
\label{fig:monomer}
\end{figure}

At this level of accuracy of the electronic structure studies, we show
   in Fig.~\ref{fig:monomer} the fully optimized geometrical
   configuration of LiBC for different spin multiplicities.
Table~\ref{tab:bonds} records the equilibrium bond lengths for Li--Li,
   Li--C, Li--B, and B--C, for some of the clusters considered in this
   study.

\begin{figure}[t]
\centering
\includegraphics[height=0.7\columnwidth,angle=-90]{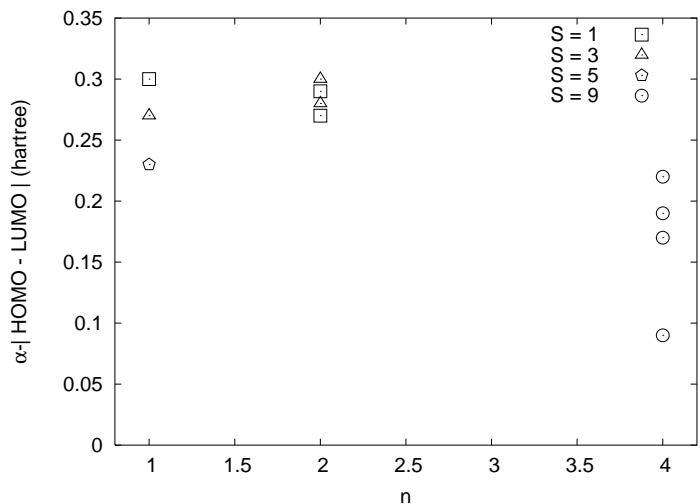}
\caption{HOMO-LUMO gap in the (LiBC)$_n$ clusters as a function of
   $n$.
Different symbols refer to the different spin multiplicity.
}
\label{fig:HL}
\end{figure}

One of the focal points of the present letter is the HOMO-LUMO energy gap
   $\epsilon_\HL$ of the various clusters considered.
Therefore in Fig.~\ref{fig:HL} for $n=1$ this gap is recorded for the
   three spin multiplicities studied.
There is not a huge spread of $\epsilon_\HL$ for $n=1$ with spin
   multiplicity, but the smallest gap is when this quantity is $\sim
   5$.
The corresponding total Hartree-Fock energies are given in
   Tab.~\ref{tab:properties} for the three different spin
   multiplicities.

\begin{figure}[t]
\centering
\begin{tabular}{ccc}
\includegraphics[width=0.3\columnwidth]{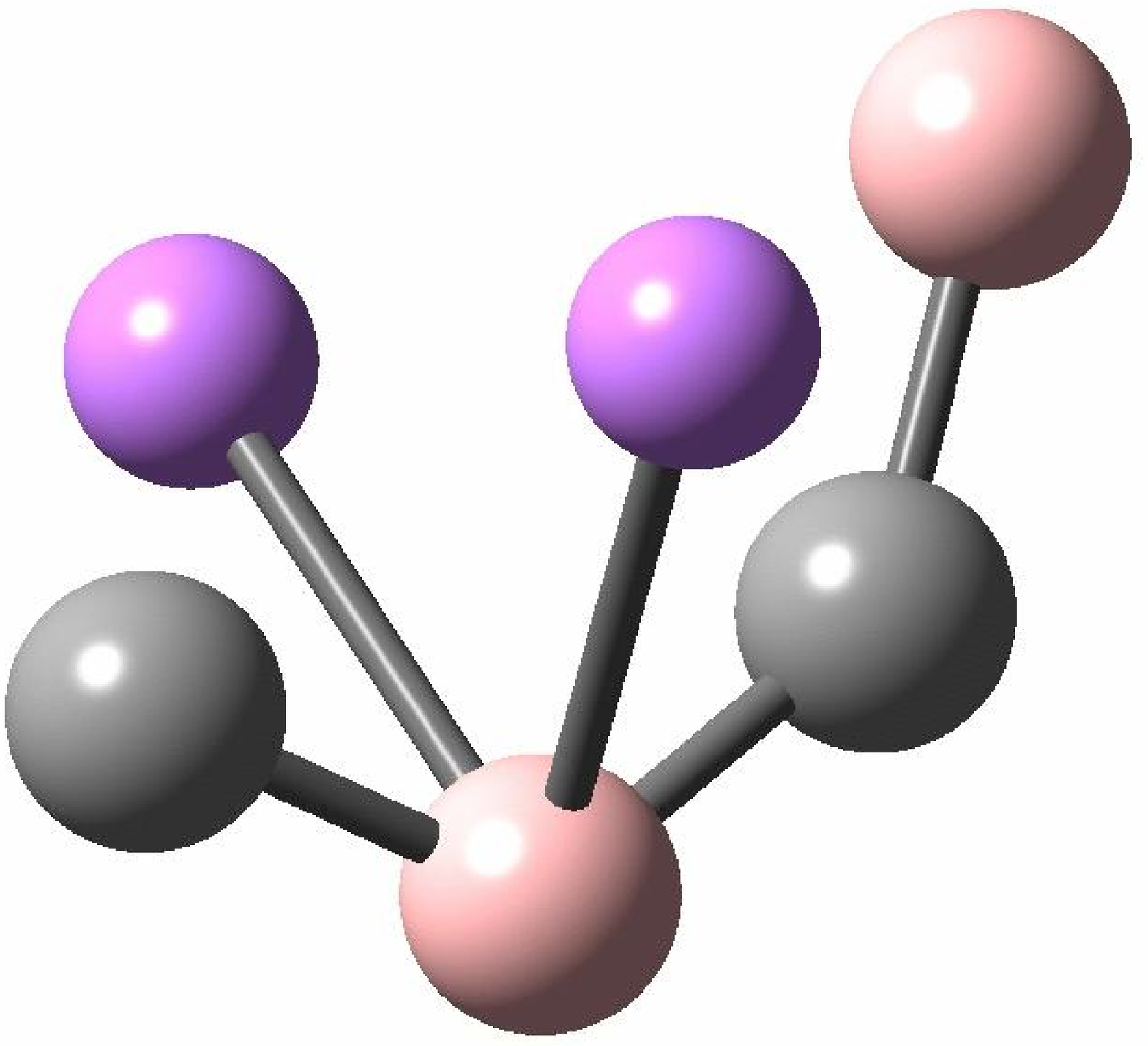}
   &
\includegraphics[width=0.3\columnwidth]{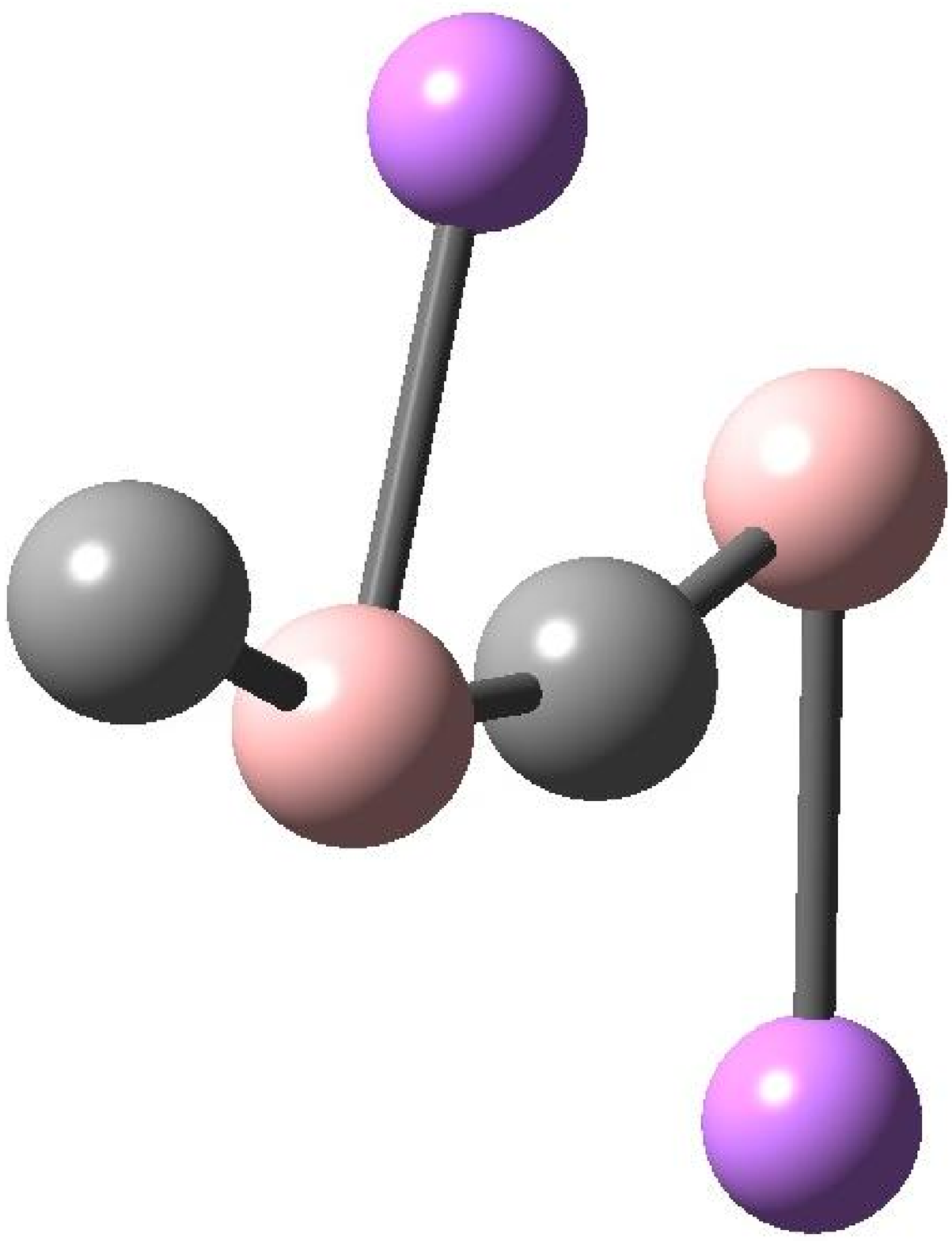}
   \\
$S=1$ & $S=1$ (trans)\\
\includegraphics[width=0.3\columnwidth]{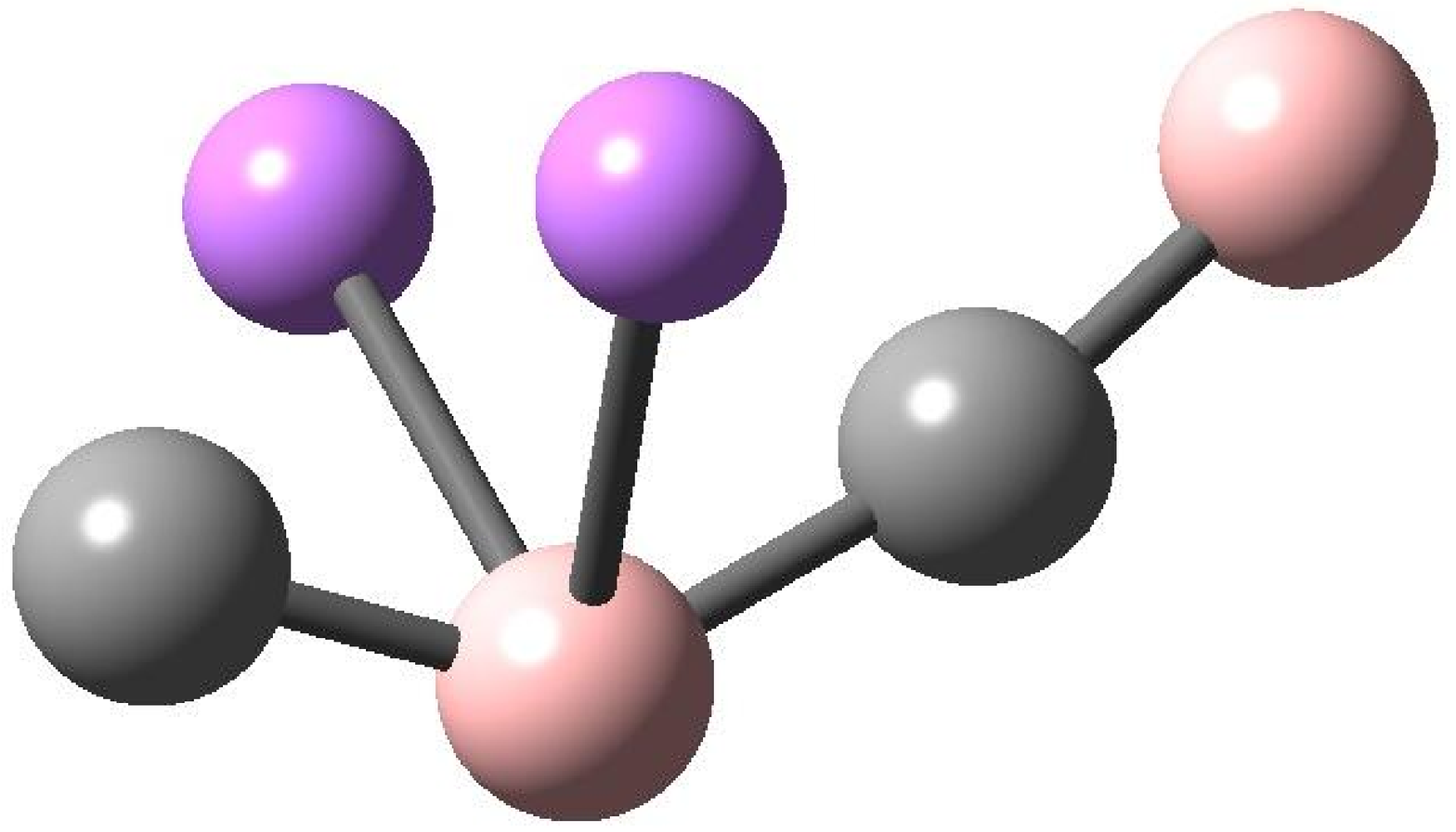}
   &
\includegraphics[width=0.3\columnwidth]{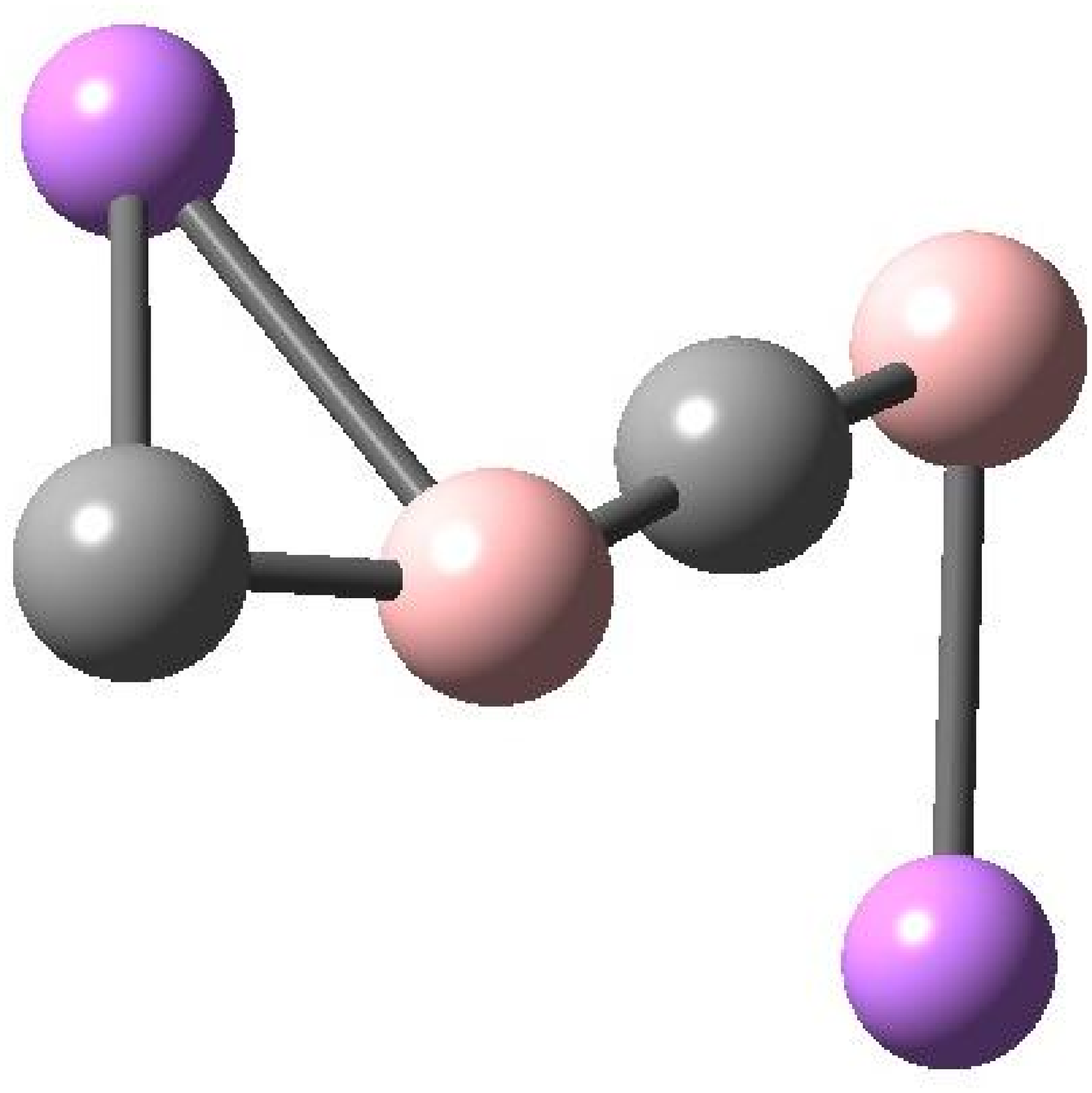}
   \\
$S=3$ & $S=3$ (trans)
\end{tabular}
\caption{{\sl (a), top row:}
Fully optimized configurations of the two isomer singlet
   (LiBC)$_2$ clusters.
{\sl (b), bottom row:}
Fully optimized configurations of the two isomer triplet
   (LiBC)$_2$ clusters.
} 
\label{fig:dimer}
\end{figure}

Turning to the case $n=2$, two geometries are shown in
   Figs.~\ref{fig:dimer}a and \ref{fig:dimer}b, with the relevant
   equilibrium bond lengths recorded in Tab.~\ref{tab:bonds}.
The `trans' configuration of Fig.~\ref{fig:dimer}, with multiplicity
   3, has the lowest energy at the present level of approximation.
The corresponding energy gaps for the two configurations are seen not
   to be very different, with also a relatively weak dependence on
   spin multiplicity.

\begin{figure}[t]
\centering
\begin{tabular}{ccc}
\includegraphics[width=0.3\columnwidth]{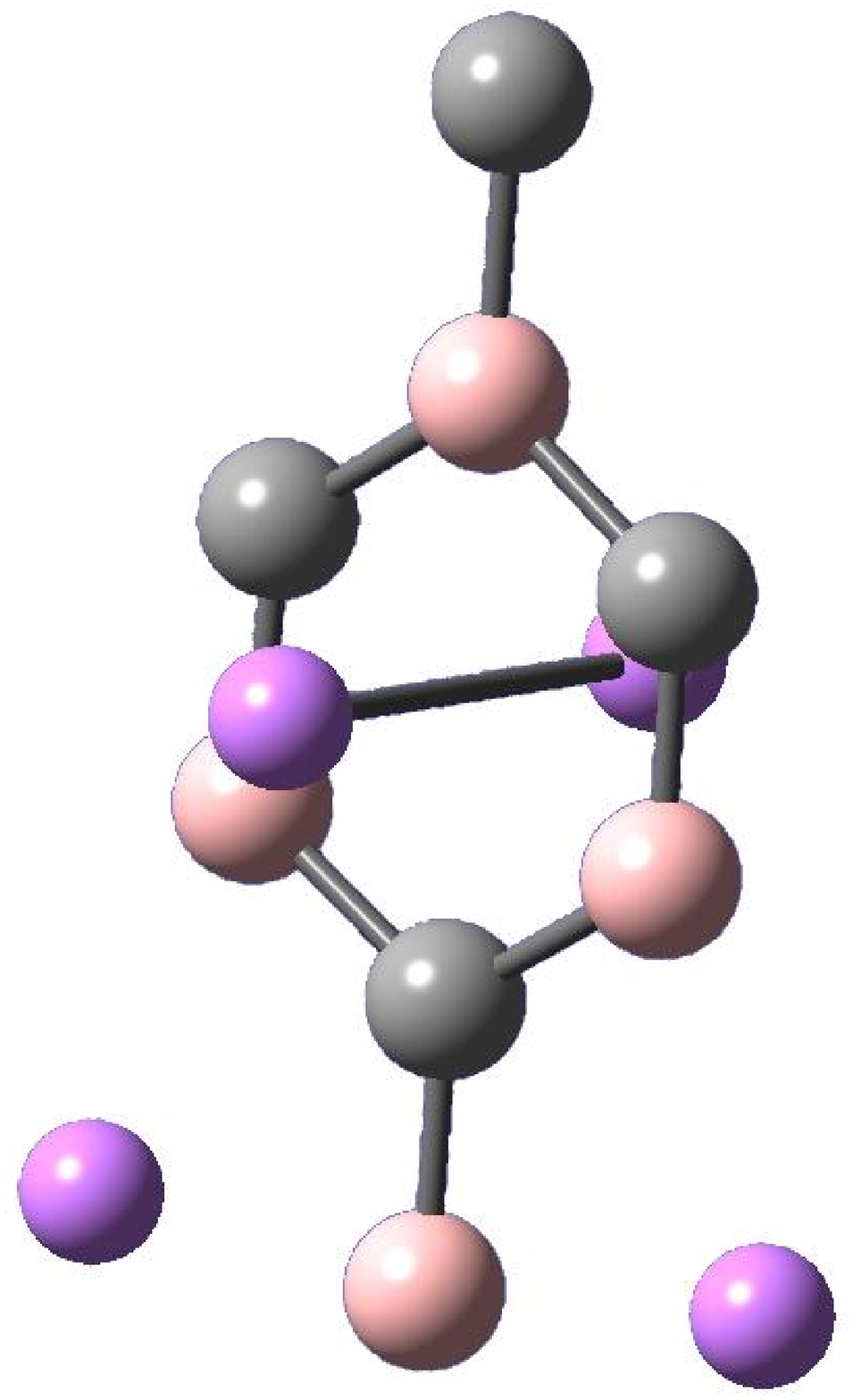}
   &
\includegraphics[width=0.3\columnwidth]{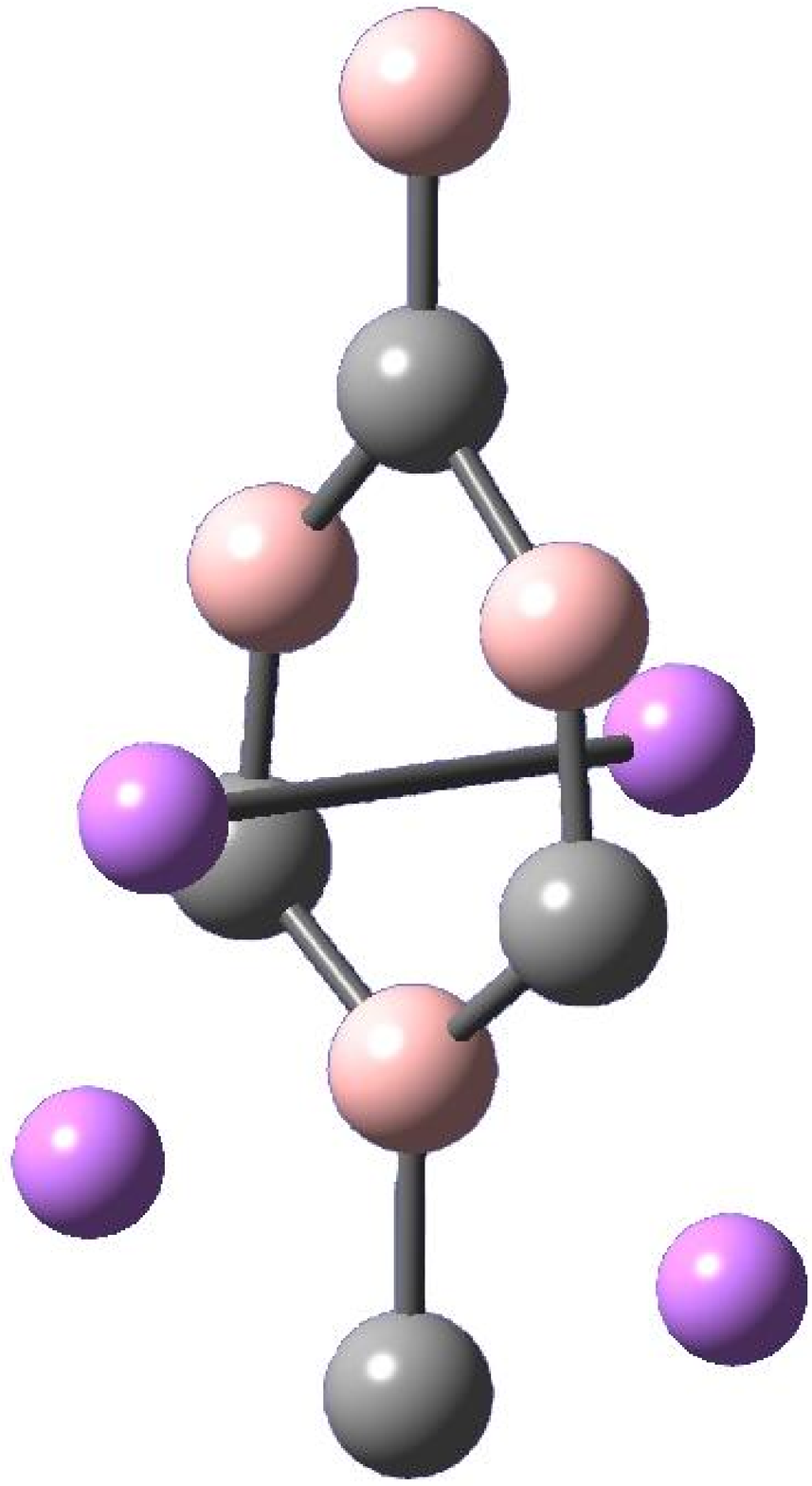}
   &
\includegraphics[width=0.3\columnwidth]{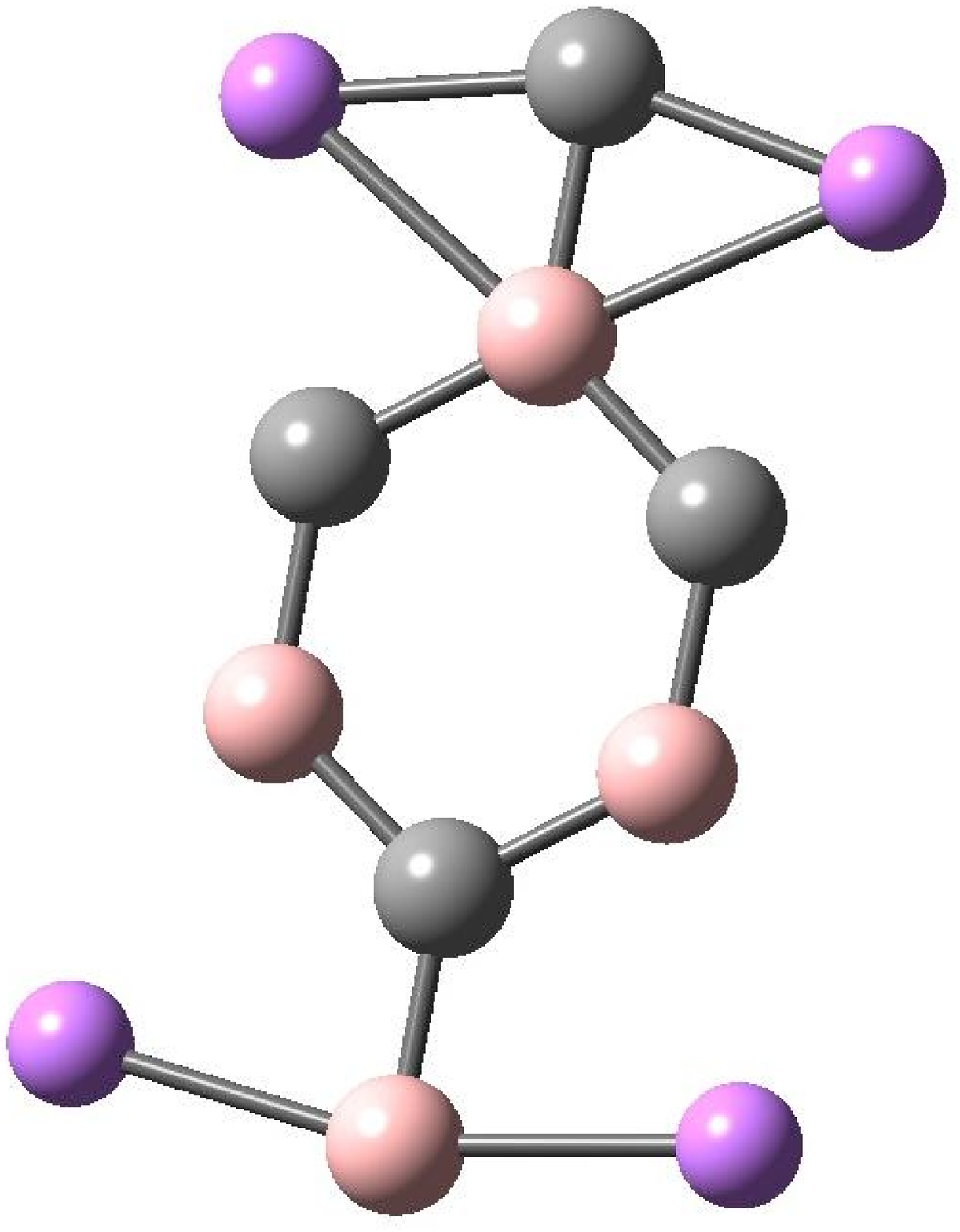}
   \\
$S=9$, C$_{2v}$ (1) & $S=9$, C$_{2v}$ (2) & $S=9$, C$_{2v}$ (3) \\
   &
\includegraphics[width=0.3\columnwidth]{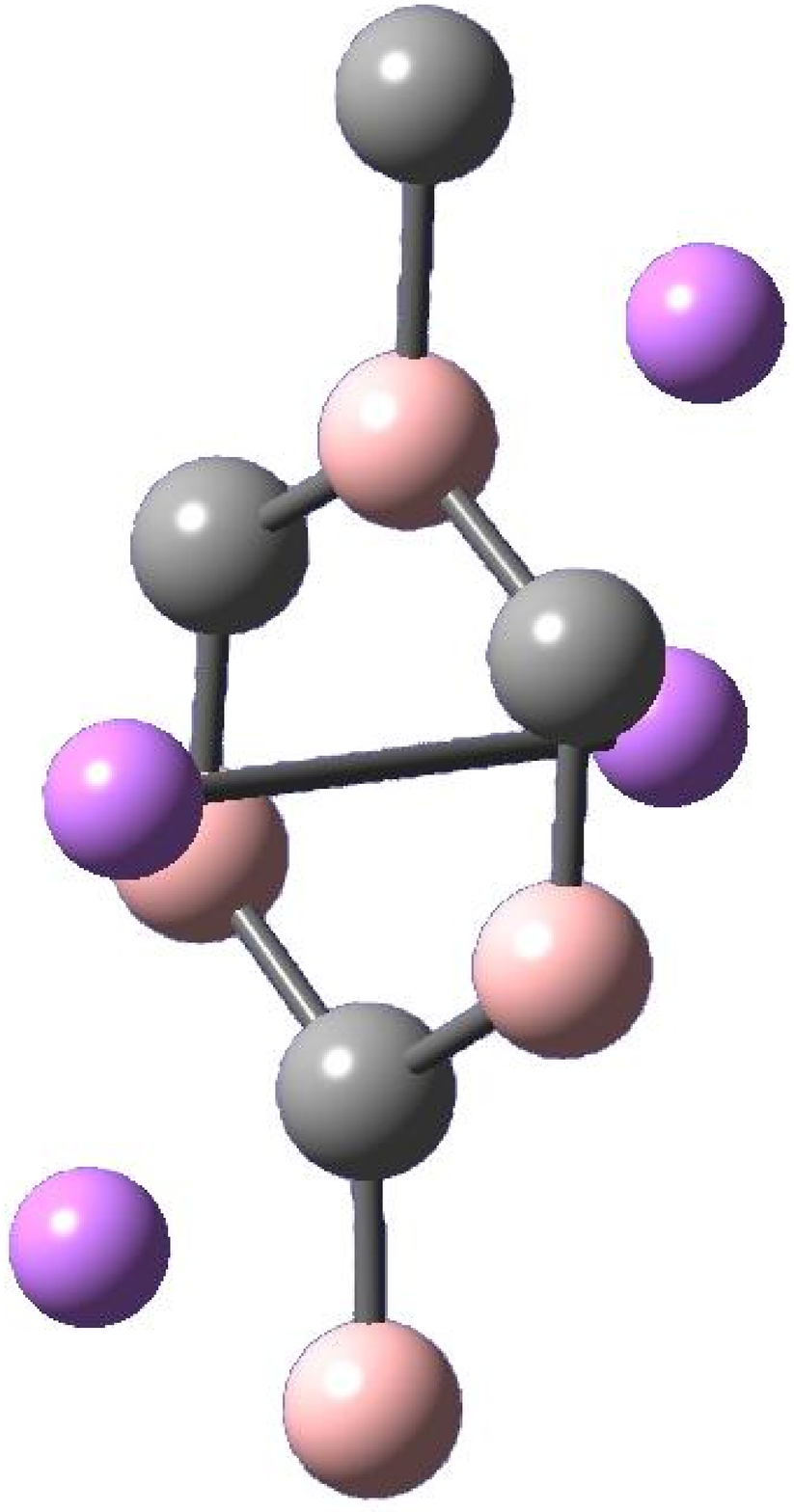}
   & \\
 & $S=9$, C$_1$ & \\
\end{tabular}
\caption{Fully optimized configurations of the four isomer quadruplet
   (LiBC)$_4$ clusters with C$_{2v}$ symmetry (top row) and C$_1$
   symmetry (bottom).
Some of the frequencies of the C$_{2v}$ (3) are imaginary, the 
   other three isomers shown being stable.
} 
\label{fig:tetramer}
\end{figure}

The largest cluster studied here corresponds to $n=4$.
Fig.~\ref{fig:tetramer} shows the fully optimized configurations of
   the four isomer quadruplet (LiBC)$_4$ clusters.
The top row of this Figure has C$_{2v}$ symmetry, whereas the lower
   Figure symmetry is C$_1$.
Studying stability via the normal mode vibrational frequencies, we
   find that some frequencies of the extreme right symmetry cluster in
   the top row of Fig.~\ref{fig:tetramer} are imaginary, all the other
   isomers presented here being stable in this context.

It is appropriate at this point to briefly discuss the geometry of the
   (LiBC)$_4$ clusters in relation to the crystalline solid form.
The main point to emphasize is that in Fig.~\ref{fig:tetramer} the
   three configurations for which the vibrational frequencies are all
   real each contain a six-membered ring, consisting of alternating
   boron and carbon atoms.
However, orthogonal to the ring, each configuration has a Li dimer
   passing through the centre of the ring.
Turning to the solid-state structure, Fig.~1 of Ref.~\cite{Rosner:02}
   contains two such BC hexagons again, which are intercalated by Li
   layers.
It should be stressed that adjacent hexagons in the solid have no
   direct B--B or C--C bonds.
One discerns a Li--Li axis passing through the centre of the BC
   hexagons, establishing thereby close contact with the cluster
   structures already discussed.
From Table~\ref{tab:bonds}, one indeed finds that the Li--Li distances
   range between 3.30 and 3.60~\AA, to be compared and contrasted with
   the interlayer Li--Li distance in solid LiBC, which is of
   3.529~\AA{} \cite{Rosner:02}.

\begin{figure}[t]
\centering
\includegraphics[width=0.8\columnwidth]{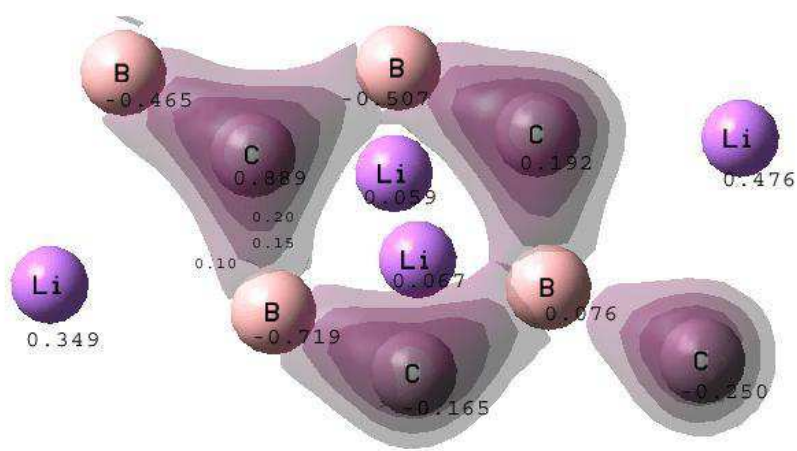}
\caption{Isosurfaces of the valence electron density for the
   (LiBC)$_4$ cluster with C$_1$ symmetry in
   Fig.~\protect\ref{fig:tetramer}.
Each isosurface is labeled with the relative value of the electron
   density ($0.10-0.20$), while each atom is labeled with its
   Mulliken atomic charge.
}
\label{fig:Mulliken}
\end{figure}

Another analogy with the solid phase is provided by the electronic
   charge distribution in the largest clusters examined in this study.
Fig.~\ref{fig:Mulliken} therefore shows the Mulliken density charge
   isosurfaces of the valence electrons, for the (LiBC)$_4$ cluster
   with C$_1$ symmetry (Fig.~\ref{fig:tetramer}).
One may detect the formation of a ring of electronic charge piling
   between the C and B atoms, which preludes to the $\sigma$ band in
   solid LiBC \cite{Rosner:02}.

Fig.~\ref{fig:HL} shows both a substantial lowering of the energy gap
   $\epsilon_\HL$ with increase in cluster size as well as a
   substantial spread in the values obtained for spin multiplicity 9
   for the four geometries depicted in Fig.~\ref{fig:tetramer}.
It is also of interest to note that the dimer (LiBC)$_2$ binding energy
   measured relative to the energy of the two separated LiBC units is
   found from Table~\ref{tab:properties} to be $\sim 4$~eV.

In conclusion, by using quantum chemical techniques, we have studied
   the \emph{Aufbau} of crystalline LiBC by gathering individual LiBC
   units into small (LiBC)$_n$ clusters.
In particular, we focussed on the structure optimization and the
   computation of selected electronic properties, such the HOMO-LUMO
   gap as a function of $n$ and the Mulliken charge density
   distribution of the largest cluster considered here ($n=4$), which
   may provide insight into the salient properties of crystalline LiBC.
Already for $n=4$ we can recognize the formation of alternating BC
   rings, with significant overlap of the valence electrons along the ring,
   and comparably less electron delocalization in the direction
   perpendicular to the ring, along which a Li--Li dimer is favoured
   to align.
This is reminiscent of the bilayered structure of crystalline LiBC,
   and of its two-band electronic character, whose relevance for
   superconductivity is well-known.

\bibliographystyle{elsart-num}
\bibliography{a,b,c,d,e,f,g,h,i,j,k,l,m,n,o,p,q,r,s,t,u,v,w,x,y,z,zzproceedings,Angilella}

\subsection*{Acknowledgements}

GGNA wishes to thank the Department of Physics, University of Antwerp,
   where this work was brought to completion, for warm hospitality and
   for the stimulating environment.
NHM acknowledges that his contribution to this study was made during a
   visit to the University of Catania.
He wishes to thank the Department of Physics and Astronomy for
   generous support.

\newpage

\begin{table}[t]
\caption{Various equilibrium bond lengths (in \AA{}) for some of the clusters considered
   in this study.} 
\label{tab:bonds}
\centering
\begin{tiny}
\begin{tabular}{lllllllllll}
\hline
\hline
 & LiBC & LiBC & LiBC & (LiBC)$_2$ & (LiBC)$_2$ & (LiBC)$_2$ &
   (LiBC)$_2$ & (LiBC)$_4$ & (LiBC)$_4$ & (LiBC)$_4$ \\
 & & & & & trans & & trans & C$_{2v}$ (1) & C$_{2v}$ (2) & C$_1$ \\
\hline
$S$ & 1 & 3 & 5 & 1 & 1 & 3 & 3 & 9 & 9 & 9 \\ 
\hline
Li--Li & & & & 3.122282 & 4.245581 & 3.0483 & 4.253679 & 3.353511 & 3.559604 &
3.300477 \\
									      
Li--C  & 1.9624 & 1.9103 & 1.9078 & 2.26   & 2.1502 & 2.298  & 2.4106 & 2.343600 & 2.646053 & 2.297274  \\
& & &                             & 2.2568 & 2.1433 & 2.2981 & 2.1345 & 2.215844 & 2.22583 & 2.266067\\  
& & & &				    2.4572 & 2.4293 & 2.6453 & 2.0751 & 2.215844& 2.22583 & 2.218518\\  
& & & &				    2.1353  &   4.6509	&  2.1503 &    4.1552
&2.3436   &2.646053 &2.292753 \\ 
& & & & & & & & 2.215844 & 2.22583  & 2.265326 \\ 
& & & & & & & & 2.215844 & 2.22583  & 2.223154  \\
& & & & & & & & 2.522509 & 2.02507  & 2.316641\\
& & & & & & & & 2.522509 & 2.02507  & 1.940265\\
& & & & & & & & & 2.476577 & 2.217949 \\
& & & & & & & &	& 2.476577 & \\
									      
Li--B  & 1.9624 & 3.289  & 2.2398 & 2.4814 & 2.574  & 3.0006 & 3.3345 &	2.146662
&2.261255 &2.213358\\
& & & &				    2.4838 & 2.1849 & 3.0007 & 2.1358 &	2.146662
&2.29926  &2.274297\\
& & & &				    2.2551 & 2.1058 & 2.1078 & 2.1539 &
2.268792& 2.261255 &2.260106\\
& & & &				    2.2543 & 3.2111 & 2.1078 & 2.8393 & 2.289784& 2.261255 &2.210945\\
& & & & & & & & 2.268792 &2.29926  &2.274707\\
& & & & & & & & 2.268792 &2.261255 &2.260612\\
& & & & & & & & 2.289784 &2.45892  &2.211303\\
& & & & & & & & 2.268792 &2.45892  &2.157141\\
& & & & & & & & 2.329519 &	&2.331481\\
& & & & & & & & 2.329519 & & \\
		    
B--C     & 1.3944   & 1.3787   & 1.6026   & 1.4185 & 1.4066 & 1.3508 & 1.4131 &
1.530041 & 1.489521 & 1.496576\\
	 & 	    &	       &	  & 3.2822 & 4.0533 & 3.8477 & 4.1366 &
1.546234 & 1.485995 & 1.567349\\
	 & 	    &	       &	  & 1.3999 & 1.4091 & 1.4583 & 1.3942 &
1.499536 & 1.473746 & 1.484584\\
	 & 	    &	       &	  & 1.4186 & 1.4496 & 1.3868 & 1.4616 &
1.534722 & 1.515207 & 1.529249\\
	 & 	    &	       &	  &	   &	    &	     &        &
1.534722 & 1.515207 & 1.516692\\
	 & 	    &	       &	  &	   &	    &	     &        &
1.571215 & 1.63323  & 1.653384\\
	 & 	    &	       &	  &	   &	    &	     &        &
1.546234 & 1.485995 & 1.620702\\
	 & 	    &	       &	  &	   &	    &	     &        &
1.499536 & 1.473746 & 1.49723\\
\hline
\hline
\end{tabular}
\end{tiny}
\end{table}

\begin{table}[t]
\caption{Calculated properties of (LiBC)$_n$ clusters. 
All energies are in hartrees.}
\label{tab:properties}
\centering
\begin{tabular}{cccr@{.}lr@{.}lr@{.}lr@{.}lr@{.}lr@{.}l}
\hline
\hline
$n$ & $S$ & & \multicolumn{2}{c}{$E_\HF$} & \multicolumn{2}{c}{$E_\HF
   /n$} & \multicolumn{2}{c}{$\alpha$-HOMO} &
   \multicolumn{2}{c}{$\alpha$-LUMO} & \multicolumn{2}{c}{$\beta$-HOMO} &
   \multicolumn{2}{c}{$\beta$-LUMO} \\
\hline
1 & 1 &  & $-69$ & 77 & $-69$ & 77 & $-0$ & 31 & $-0$ & 01 & $-0$ & 31
   & $-0$ & 01 \\
1 & 3 &  & $-69$ & 85 & $-69$ & 85 & $-0$ & 28 & $-0$ & 01 & $-0$ & 27
   & $-0$ & 01 \\
1 & 5 &  & $-69$ & 77 & $-69$ & 77 & $-0$ & 23 & $0$ & 00 & $-0$ & 34
   & $-0$ & 01 \\
& & & \multicolumn{2}{c}{} & \multicolumn{2}{c}{} &
   \multicolumn{2}{c}{} & \multicolumn{2}{c}{} & \multicolumn{2}{c}{}
   & \multicolumn{2}{c}{} \\ 
2 & 1 &  & $-139$ & 81 & $-69$ & 905 & $-0$ & 29 & $0$ & 00 & $-0$ & 29
   & $0$ & 00 \\
2 & 1 & trans & $-139$ & 81 & $-69$ & 91 & $-0$ & 28 & $-0$ & 01 & $-0$ & 28
   & $-0$ & 01 \\
2 & 3 & & $-139$ & 84 & $-69$ & 92 & $-0$ & 28 & $0$ & 00 & $-0$ & 27
   & $0$ & 00 \\
2 & 3 & trans & $-139$ & 85 & $-69$ & 92 & $-0$ & 30 & $0$ & 00 & $-0$ & 28
   & $-0$ & 01 \\
& & & \multicolumn{2}{c}{} & \multicolumn{2}{c}{} &
   \multicolumn{2}{c}{} & \multicolumn{2}{c}{} & \multicolumn{2}{c}{}
   & \multicolumn{2}{c}{} \\ 
4 & 9 & C$_{2v}$ (1) & $-279$ & 69 & $-69$ & 92 & $-0$ & 17 & $0$ & 00
   & $-0$ & 26 & $0$ & 00 \\ 
4 & 9 & C$_{2v}$ (2) & $-279$ & 69 & $-69$ & 92 & $-0$ & 11 & $-0$ &
   02 & $-0$ & 30 & $0$ & 00 \\  
4 & 9 & C$_{2v}$ (3) & $-279$ & 73 & $-69$ & 93 & $-0$ & 22 & $0$ & 00
   & $-0$ & 24 & $0$ & 00 \\  
4 & 9 & C$_1$ & $-279$ & 73 & $-69$ & 93 & $-0$ & 19 & $0$ & 00 &
   $-0$ & 25 & $0$ & 00 \\
\hline
\hline
\end{tabular}
\end{table}

\end{document}